\documentclass[superscriptaddress,article,pra,showpacs,nofootinbib,11pt]{revtex4-1}
\usepackage{amsmath,xcolor,amsfonts,amsthm,amssymb,graphicx, srcltx,hyperref}
\usepackage[normalem]{ulem} 
\usepackage{tikz}
\usepackage{acronym}

\acrodef{pdf}[PDF]{probability density function}
\newcommand{\beq}{\begin{equation}}
\newcommand{\eeq}{\end{equation}}
\newcommand{\ket} [1] {|#1\rangle}

\newcommand{\deltasec}{\delta_{\rm{sec}}}
\newcommand{\epssec}{\varepsilon_{\rm{sec}}}
\newcommand{\epscor}{\varepsilon_{\rm{cor}}}

\newcommand{\Qall}[1]{Q_{\mathbb{#1}}}
\newcommand{\Qkey}[1]{\hat{Q}_{\mathbb{#1}}}
\newcommand{\Qtol}{Q_{\rm{tol}}^{\mathbb{Z}}}
\newcommand{\Qmax}{Q_{\rm{max}}^{\mathbb{X}}}

\newcommand{\ent}{H}
\newcommand{\mutinfo}{I}
\newcommand{\iacc}{\mutinfo_{\rm{acc}}}
\newcommand{\kldist}{\mathbb{D}}
\newcommand{\expect}{\mathbb{E}}
\newcommand{\prob}{\mathbb{P}}
\newcommand{\key}{\mathbf{S}}
\newcommand{\ukey}{\mathbf{U}_{\bf S}}
\newcommand{\pabort}{p_{\rm{abort}}}

\begin{document}

\title{Experimental quantum key distribution with finite-key security analysis for noisy channels}
\author{Davide Bacco}
\affiliation{Department of Information Engineering, University of Padova, via Gradenigo 6/B, 35131 Padova, Italy}
\author{Matteo Canale}
\affiliation{Department of Information Engineering, University of Padova, via Gradenigo 6/B, 35131 Padova, Italy}
\author{Nicola Laurenti}
\affiliation{Department of Information Engineering, University of Padova, via Gradenigo 6/B, 35131 Padova, Italy}
\author{Giuseppe Vallone}
\affiliation{Department of Information Engineering, University of Padova, via Gradenigo 6/B, 35131 Padova, Italy}
\author{Paolo Villoresi}
\email {paolo.villoresi@dei.unipd.it}
\affiliation{Department of Information Engineering, University of Padova, via Gradenigo 6/B, 35131 Padova, Italy}
\date{\today}

\begin{abstract}
In quantum key distribution implementations, each session is typically chosen long enough so that the secret key rate approaches its 
asymptotic limit. However, this choice may be constrained by the physical scenario, 
as in the perspective use with satellites, where the passage of one terminal over the other is restricted to a few minutes.
Here we demonstrate experimentally the extraction of secure keys leveraging an optimal design of the prepare-and-measure scheme, 
according to recent finite-key theoretical tight-bounds. 
The experiment is performed in different channel conditions, and assuming two distinct attack models: individual attacks, 
or general quantum attacks.
The request on the number of exchanged qubits is then obtained as a function of the key size and of the ambient 
quantum bit error rate. The results indicate that viable conditions for effective symmetric, and even one-time-pad, cryptography are achievable.
\end{abstract}

\maketitle

\section{Introduction}
Quantum key distribution (QKD) is a technique for sharing a  random secret key {by means of a quantum link} between two distant partners, traditionally called Alice and Bob. For this purpose, an optical link is established with Alice acting as the sender and Bob as the receiver in a 
prepare-and-measure scenario, or with both receiving a signal from an intermediate source~\cite{scar09rmp}.
The secret key that is obtained may be used in any symmetric cryptographic algorithm including the {\it one-time pad} encryption introduced by Vernam~\cite{vern26jai} or computationally secure ciphers {such} as AES.

QKD may be considered the first successful example of a quantum information protocol that reached the everyday applications. Indeed,  commercial devices communicating via optical cables are already operated worldwide. The perspective use in free space is also considered very attractive. This use includes terrestrial links, in the case that it is not possible to use optical cables, or in the case that either terminal is moving, including the very relevant case of key exchange with orbiting terminals, that is, satellite QKD. This extension of the QKD application has been fostered for years, being included in the major Quantum Information Roadmaps \cite{eu-roadmap,jp-roadmap,USA-roadmap}, and has been the subject of several feasibility studies \cite{nord02spie,aspe03ieee, peng05prl, bona09njp, toma11asr, meye11pra,capr12prl}.

However, the intrinsic difficulties in its realization  allowed only the experimental demonstration of the single photon exchange with an orbiting terminal  \cite{vill08njp}. 
Moreover, in free-space links the gathering of light from the background is much more pronounced than for optical fibers. At the same time, in the case of long distance terrestrial links or space to ground links, signal attenuation is typically greater by at least three orders of magnitude. As a consequence, strong noise overimposed to an attenuated signal results in a poor signal to noise ratio (SNR) and in an increased quantum bit error rate in the sifted key. 

The experimental investigation of such limit is therefore of crucial interest, in order to open the way to direct experiments in the free-space QKD, and the recent result on finite-key bounds by Tomamichel {\it et al.} provides the necessary theoretical framework \cite{toma12nco}. As the final goal of this work we aim to prove experimentally the bound for the number of exchanged raw key bits that is necessary to extract a secret key of desired length. This is the recipe needed to design the terminal dimension and performance in practical applications. 

Any QKD protocol consists of a physical quantum communication layer and a  post-processing layer, in which,  by using a classical communication channel, the secret key is  extracted from the raw data  shared by the two terminals: first {the raw data is sifted in order to distill maximally correlated data between Alice and Bob},   {then}  an information reconciliation protocol is performed in order to correct the errors between the two users and finally a privacy amplification algorithm is {used} in order to ensure the secrecy of the final key. 

A crucial parameter is the so called {secure key rate}, i.e., the {ratio of the number} of secret bits that can be extracted to the number of correlated, or raw, bits obtained in the quantum layer of the protocol. According to standard QKD unconditional security proofs,  the secret key rate is upper-bounded by the asymptotic limit which is achievable in the limit of infinitely long keys (see for instance~\cite{scar09rmp}), {with the use of shorter blocks leading to lower key rates}.
{However, in QKD implementations, the length of processed blocks is chosen as a trade-off between link duration constraints and memory resources on one side and efficiency (in terms of secret key rate) on the other. 
This trade-off usually results in long blocks, of at least a million sifted bits. However, in some scenarios such a choice may rather be constrained by the physical channel, {as in the perspective use with satellites, where the passage of the orbiting terminal over the 
ground station is restricted to a few minutes in the case of low-Earth-orbit (LEO) satellite \cite{vill08njp,bona09njp}} or to a fraction of one hour for the medium-Earth-orbit (MEO) ones  \cite{toma11asr}.}
{Hence}, for practical use of QKD in cryptography, it is of crucial importance to develop and test methods that give the {achievable} secure key rates in the {bounded-key-length} scenario, since the number of exchanged bits between the two parties is always finite.
In the last years, great efforts from the quantum communication community were  directed to this subject, due to its relevance for a number of application scenarios~\cite{hase07qph, scar08prl,rice09qph,rose09njp,cai09njp,scar10proc,abru11pra}.
We would like to underline that all previous published experimental work on finite-size key security were based on a far more inefficient bound as compared with the one obtained in Tomamichel {\it et al.}, Ref. \cite{toma12nco}.

In this work, we study the security and the generation rate of a protocol for key exchange in the finite-key regime and in presence of noise, whose value is experimentally varied up to the top limit. The security is assessed with reference to a recently introduced theoretical result \cite{toma12nco},
for which ``almost tight bounds on the minimum value" of exchanged qubits ``required to achieve a given level of security" were obtained
\cite{toma12nco}, as well as for a realistic bound described below.
{In particular, by leveraging the optimal design of the prepare-and-measure scheme complying with the above mentioned tight theoretical bounds, we evaluate how the secret key rate scales in different channel conditions, depending on the protocol parameters. We consider two possible attack models, referring to two different levels of secrecy: \emph{pragmatic secrecy}, which ensures resiliency against individual attacks, and \emph{general secrecy}, which ensures resiliency against the most general quantum attacks.}

\section{Results}
\subsection{Protocol for quantum key distribution.} 
We will adopt here the protocol described in \cite{toma12nco}, a derivation of the well known BB84 protocol~\cite{BB84}.
 According to this protocol, one of the two bases is used to encode the raw key bits while the other basis is used  to test the channel {for the presence of the eavesdropper}~\cite{lo05jcrc}. Moreover, the two bases are selected by Alice and Bob in the preparation of the qubits and in their measure, respectively, with non equal probabilities, unlike the standard BB84. 

Let us describe in more detail the quantum communication part of the QKD protocol used in the present experiment, characterized by the sifted key length $n$ and the number of bits used for parameter estimation $k$; both parameters can be chosen according to the required secret key length and channel conditions as described below.
Alice prepares and sends to Bob quantum states encoded by means of photon polarization. She can choose between two bases, 
$\mathbb{X} = \lbrace \ket{H}, \ket{V} \rbrace$ and  $\mathbb{Z} = \lbrace \ket{+}, \ket{-}\rbrace$ with 
$\ket{\pm}\equiv ( \ket{H} \pm \ket{V})/\sqrt2$. For each basis, the first state represents the bit 0 and the second state the bit 1.
Alice sends to Bob the raw key (namely a sequence of uniformly random bits) by randomly and asymmetrically encoding the bits with one of the two bases: with probability $p_{\mathbb{X}}=\frac{1}{1+\sqrt{k/n}}$ she encodes the bits in the $\mathbb{X}$ basis and with probability $p_{\mathbb{Z}}=1-p_{\mathbb{X}}$ she encodes the bits in the $\mathbb{Z}$ basis. 
Bob measures the photons by randomly choosing a basis, $\mathbb{X}$ or $\mathbb{Z}$, with the same probabilities $p_{\mathbb{X}}$ and $p_{\mathbb{Z}}$.

Alice and Bob broadcast their bases choices over the classical channel and Bob also communicates when he received the photons; bits corresponding to non received photons are discarded. 
Otherwise, when Alice and Bob have both chosen the same basis (it happens with probability $p^2_{\mathbb X}$ for the $\mathbb{X}$ basis and with probability $p^2_{\mathbb Z}$ for the $\mathbb{Z}$ basis) they store the respective bits, while, when they have chosen different bases, their bits are discarded.
The protocol repeats the quantum communication as long as either the number of $\mathbb{X}$  bits is lower than $n$ or 
the number of $\mathbb{Z}$  bits is lower than $k$.
In order to obtain the final sifted keys, Alice and Bob keep the same $n$ bits, randomly chosen, from the $\mathbb{X}$ bits
to form the sifted key strings $\mathbf{X}=\{x_i\}$ and $\mathbf{X'}=\{x'_i\}$.
Similarly they choose $k$ random bits from the $\mathbb{Z}$ bits to obtain the parameter estimation strings 
$\mathbf{Z}=\{z_i\}$ and $\mathbf{Z'}=\{z'_i\}$. {Differently from \cite{toma12nco}, we defined the sifted key as
 $\mathbf{X}$ and not as the union set of $\mathbf{X}$ and $\mathbf{Z}$}. The $\mathbb X$ bits will be used to build the final secret key and the expected number of errors between $\mathbf{X}$ and $\mathbf{X'}$ is the crucial parameter in the design of 
 the information reconciliation protocol. The $\mathbb Z$ bits will be used to test the presence of the eavesdropper and the number of errors between $\mathbf{Z}$ and $\mathbf{Z'}$ is used for dimensioning the privacy amplification procedure.  
Note that the probabilities $p_{\mathbb{X}}$ and $p_{\mathbb{Z}}$ are chosen to satisfy 
${p^2_{\mathbb Z}}/{p^2_{\mathbb X}}={k}/{n}$ in order to minimize the number of exchanged photons before the quantum communication is stopped.

After the quantum transmission and the sifting of the raw data, four subsequent tasks take place: parameters estimation, information reconciliation, error verification and privacy amplification. The first task, parameters estimation, is required to measure the quantum bit error rate (QBER) on the $\mathbb{Z}$ basis, $\Qall{Z}$. 

Furthermore, we assume that the quantum channel is stable, i.e., that QBER on the $\mathbb{X}$-basis, $\Qall{X}$, is constant in time (note that, in general, $\Qall{X} \neq \Qall{Z}$). If $\Qall{X}$ increases (for instance because an attacker is tampering with the channel), then the information reconciliation will fail. The failure will be detected during the error verification phase, and the protocol will abort. 
On the other hand, the empirical QBER in the $\mathbb{Z}$ basis is dynamically computed at each protocol run as 
$\Qkey{Z} = (\sum_{i = 1}^k z_i \oplus z'_i)/k$, to check for the presence of an eavesdropper. The protocol aborts if $\Qkey{Z} > \Qtol$, where $\Qtol$ is a given channel error tolerance on the $\mathbb{Z}$ basis which has been determined a priori based on the expected behavior of the quantum channel and the required level of security.

Information reconciliation allows Bob to compute an estimate $\mathbf{\hat{X}}$ of $\mathbf{X}$ 
{by revealing $L_{\rm{EC}}$ bits ($L_{\rm{EC}}$ represents the classical information leakage).} 
We define $P_{\rm{fail}}$ as the upper bound to the probability of a reconciliation failure and $\epscor$ as the upper bound to the probability that $\mathbf{\hat{X}}$ differs from $\mathbf{X}$.
{We fixed a threshold $Q_{\rm{max}}^{\mathbb{X}}$ such that the empirical QBER $\Qkey{X}$ in the sifted key is higher than $Q_{\rm{max}}^{\mathbb{X}}$ with probability less than $P_{\rm{fail}}/2$. }
For details on {the chosen} information reconciliation, error verification and privacy amplification {mechanisms}, see the Methods section.

\subsection{General and pragmatic secrecy}
As introduced above, in this work we consider two possible attacker models, which in turn entail two different notions of secrecy, which we call {\it general} and {\it pragmatic}, respectively. General secrecy, { as  defined in} \cite{toma12nco}, requires that the final shared keys are secret with respect to the most general quantum attacks, and it is based on the secrecy criterion provided in \cite{kon07prl}. 
{We say that} the distilled key $\key$  is $\epssec$-GS (general secret) if for any attack strategy
\begin{equation}
\min_{\sigma_{\rm{E}}} \frac{1}{2} \lVert \rho_{\rm{SE}}-\omega_{\rm{S}} \otimes \sigma_{\rm{E}} \rVert_1 \leq \frac{\epssec}{(1 - \pabort)}, \label{eq:GS_sec}
\end{equation}

\noindent being $\lVert \rho \rVert_1 = \rm{Tr}\sqrt{\rho\rho^\dag}$, $\pabort$ the probability that the protocol aborts, $\rho_{\rm{SE}}$ the quantum state which describes the correlation between Alice's classical key $\key$ and the eavesdropper, $\omega_{\rm{S}}$ the fully mixed state on $\key$, and $\sigma_{\rm{E}}$ a generic quantum state on the eavesdropper's Hilbert space. 
Then, if the bases $\mathbb X$ and $\mathbb Z$ are chosen as described above and assuming that Alice uses an ideal single photon source, 
the authors of \cite{toma12nco} show that an $\epssec$-GS key can be extracted out of the reconciled key, with length
\begin{equation}\label{eq:l_gs}
\ell \leq n(1 - \widetilde h_2(\Qtol+ \mu)) - L_{\rm{EC}}-\log_2\frac{2P_{\rm{fail}}}{\epssec^2\epscor}
\end{equation}
where $\mu = \sqrt{\frac{n+k}{nk}\frac{k+1}{k}\ln{\frac{2}{\epssec}}}$, 
$h_2(x)=-x\log_2x-(1-x)\log_2(1-x)$ is the binary Shannon entropy function, $\widetilde h_2(x)=h_2(x)$ for $0\leq x\leq0.5$ and
 $\widetilde h_2(x)=1$ for  $x>0.5$.

On the other hand, pragmatic secrecy \cite{cana11isa} ensures that the final key is secret with respect to intercept-and-resend (IS) attacks \cite{huttner92}, i.e., a specific class of selective individual attacks, which, however, represents the most realistic and feasible attack strategy based on the experimental technology nowadays available: collective or more general attack models (see \cite{scar09rmp}), in fact, require ancillary qubits and quantum memories in order to be deployed.

While in a long-term perspective (more than 50 years) general security is the goal, in the near future (5-10 years), we know that an ideal IS attack is the best option that an eavesdropper can choose because the quantum memory needed for a general or coherent attack is not yet available. {In the Experimental Results subsection,} we will show that there are situations in which no key can be extracted if general security is required, while a pragmatically secure secret key can be obtained. In these cases, requiring general security, a protection far above actual possibilities of an eavesdropper, prevents key generation. Also, we would like to stress that pragmatic secrecy, unlike computational secrecy, offers forward security: if a key is produced today with pragmatic secrecy (without quantum memory available for Eve), the key or a message encrypted with it will be secure for any future use.

As a criterion for pragmatic secrecy, we use a bound on the classical equivocation at the eavesdropper, namely we say that the distilled key $\key$ is $\deltasec$-PS (pragmatic secret) if, for any IS attack strategy and in the case that the protocol is not aborting,
\begin{equation}
\ent(\ukey) - \ent(\key|V) \leq \frac{\deltasec}{1-\pabort} \label{eq:PSdef}
\end{equation}
being $\ukey$ the uniform key with the same length as $\key$, $V$ the classical random variable which summarizes all the information available to the eavesdropper and $\ent(\key|V)$ the equivocation (conditional entropy) of $\key$ given $V$. Note that eq. (\ref{eq:PSdef}) implies the uniformity and the security conditions
\beq
\label{eq:PSdef_bis}
\begin{cases}
 \ \ent(\key) \geq \ent(\ukey) - \frac{\deltasec}{1-\pabort} \qquad \mbox{(uniformity)}\\
 \ \iacc(\key;E) \leq \frac{\deltasec}{1-\pabort}  \qquad\qquad \ \mbox{(security)}
\end{cases}
\eeq
where the accessible information $\iacc$ is the maximum mutual information $\mutinfo(\key; V)=\ent(\key)-\ent(\key|V)$ that can be extracted from the quantum system $E$ \cite{kon07prl}. Moreover, choosing $\deltasec = \frac{2}{\ln 2}\epssec^2$ in (\ref{eq:PSdef}) implies condition (\ref{eq:GS_sec}) for non-coherent attacks (see Methods section).
{It should be noted that,} as for incoherent individual attacks, eq. \eqref{eq:PSdef} guarantees composable security, 
as the eavesdropper, without a quantum memory, cannot exploit the ``locking property'' of the accessible information (see \cite{kon07prl}).

The pragmatic security of the distilled key can be assessed through the following result, the proof of which is provided in the Methods section.

Theorem 1: 
The distilled key $\key$ is $\deltasec$-PS if
 \begin{equation} \label{eq:thm_pragmatic_secrecy}
\exists \ a \in \mathbb{N}: \quad f(a,\ell) \leq \deltasec
 \end{equation}
where 
\begin{equation}
f(a,\ell) = \ell \ \max_{q} \left[I_q(a+1,n-a) I_{1-q/2}(k(1-\Qtol),k\Qtol+1) \right] + \frac{2^{-(n_{\rm{EC}}-\ell-a)}}{\ln2}, \label{eq:fdef}
 \end{equation}
with $n_{\rm{EC}} = n - L_{\rm{EC}} - \lceil \log_2 (P_{\rm{fail}}/\epscor) \rceil$
and $I_x(a,b)$ denoting the regularized incomplete beta function \cite[section 6.6]{Abramowitz},
\beq
I_x(a,b) = \frac{B(x;a,b)}{B(1;a,b)} \ ,\ \ B(x;a,b) = \int_0^x t^{a-1}(1-t)^{b-1}\,dt.
\eeq

 Based on (\ref{eq:thm_pragmatic_secrecy}), we can therefore choose the optimal secret key length as
\begin{equation}
\label{eq:l_ps}
 \ell = \max \left\lbrace b: \min_{a} f(a,b) \leq \deltasec
 \right\rbrace
\end{equation}
Please note that, in order to  allow a comparison with the tight bound (\ref{eq:l_gs}), we have derived the secure key length in the hypothesis that Alice uses a single photon source.

Finally, given the probability $\varepsilon_{\rm{rob}}$ that the protocol aborts even if the eavesdropper is inactive \cite{toma12nco}, we can compute the final secret key rate for both general and pragmatic secrecy as
\begin{equation} \label{eq:sec_key_rate}
 r(\ell,n,k,\varepsilon_{\rm{rob}}) = (1- \varepsilon_{\rm{rob}}) \frac{\ell}{M(n,k)}
\end{equation}
where $M(n,k) = n+k+2\sqrt{nk}$ is the expected number of qubits that have to be sent until $n$ sifted key bits and $k$ parameter estimation bits are collected.

\subsection{Experimental results}

\begin{figure}
\centering 
\includegraphics[width=0.5\textwidth]{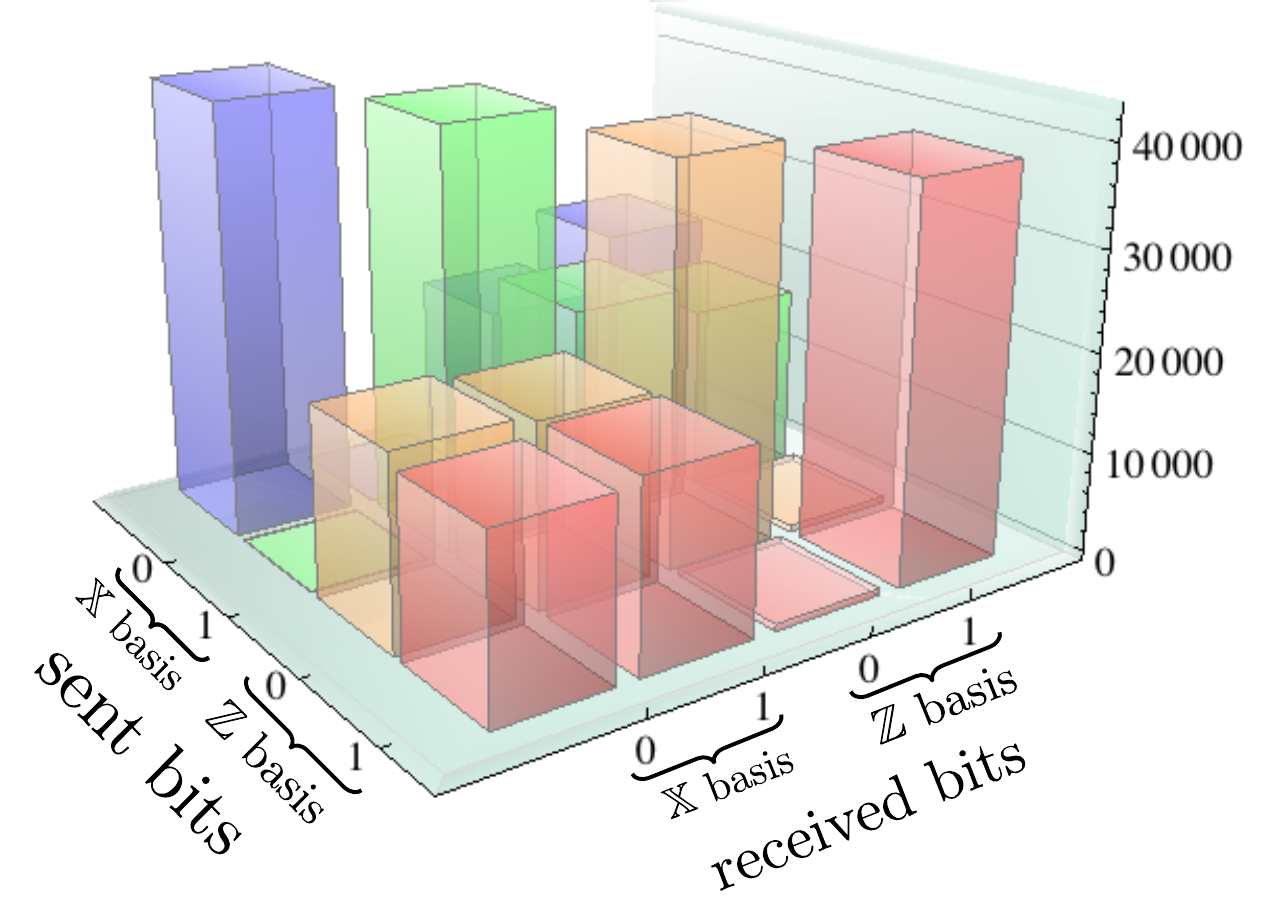} 
\caption{{\bf Experimental bits.}
Joint empirical distribution of sent and received bits, as obtained in one experiment with the best channel conditions 
(corresponding to $\Qall{X}=0.33\%$ and $\Qall{Z}=1.48\%$). The probabilities of sending and measuring in the
${\mathbb{X}}$ and ${\mathbb{Z}}$ basis were $p_{\mathbb{X}}=0.51$ and $p_{\mathbb{Z}}=0.49$, respectively.}
\label{fig:data}
\end{figure}

\begin{figure}
\centering 
\includegraphics[width=0.9\textwidth]{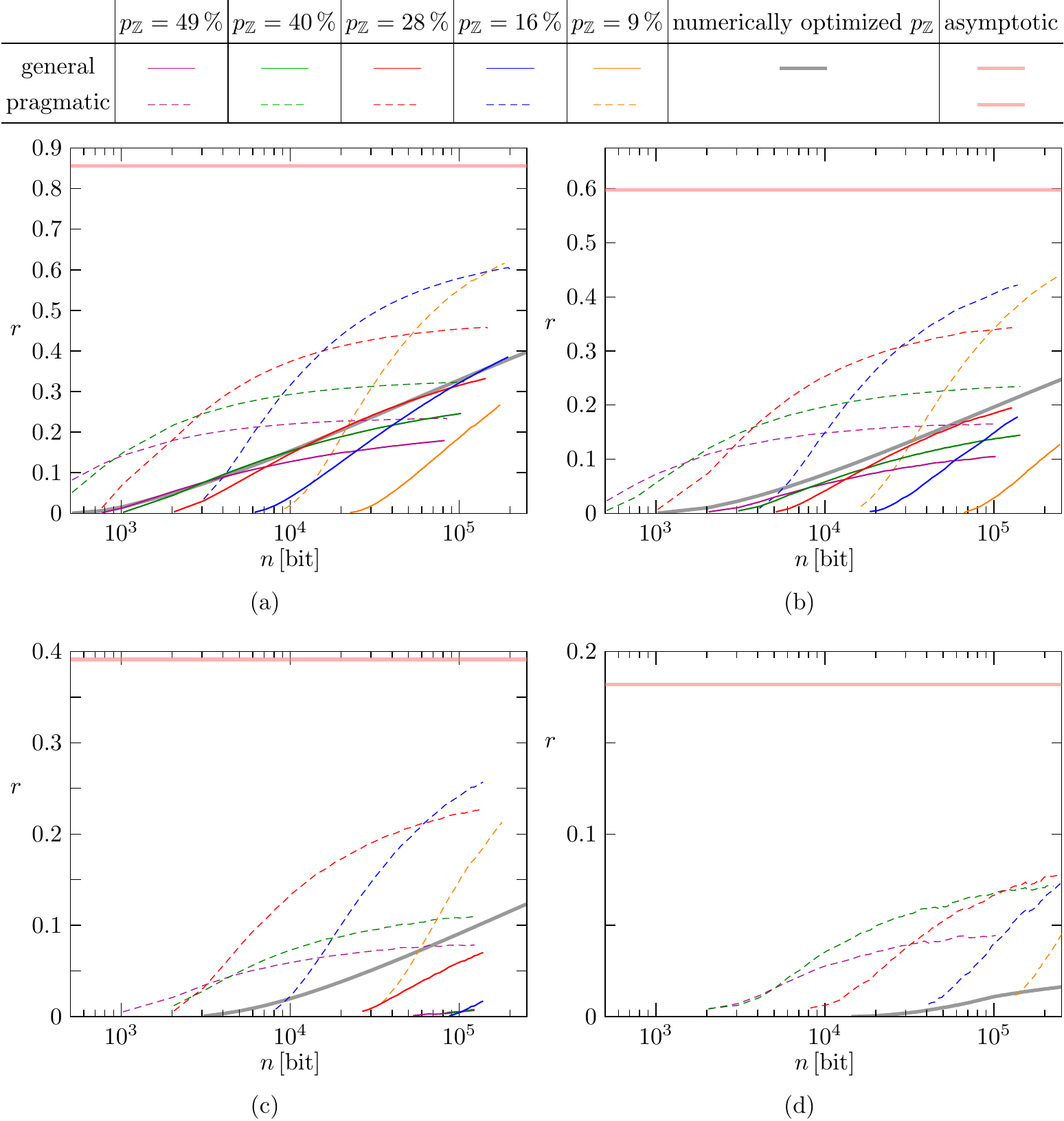} 
\caption{{\bf Experimental key rates.} Experimental secret key rates $r$ vs.\ sifted key length $n$ for different probabilities of encoding and measuring on the two bases $p_{\mathbb{Z}}$, $p_{\mathbb{X}}=1-p_{\mathbb{Z}}$ and for different channel conditions (values of the average QBERs $\Qall{X},\Qall{Z}$): 
(a) $\Qall{X} = 0.3\,\%$, $\Qall{Z} = 1.5\,\%$; (b) $\Qall{X} = 2.4\,\%$, $\Qall{Z} = 3.9\,\%$;
(c) $\Qall{X} = 4.9\,\%$, $\Qall{Z} = 6.0\,\%$; (d) $\Qall{X} = 8.3\,\%$, $\Qall{Z} = 8.1 \,\%$.
For each case we report the key rates obtained for $\epssec$-GS (solid lines) and {$\deltasec$-PS} (dashed lines) keys with $\epssec = 10^{-10}$, {$\deltasec = \frac{2}{\ln 2}\epssec^2$}, $P_{\rm fail}=10^{-3}$ and a correctness parameter $\epscor = 10^{-10}$. 
The standard deviation of experimental rates are on the order of $10^{-3}$ for both $\epssec$-GS and {$\deltasec$-PS} keys. Error bars are not reported in the plot for the sake of clarity.
For comparison, we also report the asymptotic key rate in the infinite length limit, and the $\epssec$-GS bound achievable by optimizing the probability $p_{\mathbb{Z}}$ and the thresholds $\Qtol,\Qmax$ for each value of $n$.}
\label{fig:key_rate_ber}
\end{figure}

We conducted experiments with different noisy channels yielding different values for the average QBERs $\Qall{X}$ and $\Qall{Z}$, each of them realized with different encoding probabilities $(p_{\mathbb{Z}},p_{\mathbb{X}})$.	
We varied the noise value in the channel by coupling to the receiver an external unpolarized source of suitable
intensity, that increased the background signal.
It is worth noting that by this operation we are modelling the following depolarizing channel 

\begin{equation}
 \mathcal C:\rho\rightarrow(1 -P)\rho+\frac{P}{4}\sum^3_{j=0}\sigma_j\rho\sigma_j,
\end{equation}
where $\sigma_j$ are the Pauli matrices, being $\sigma_0$ the identity and $P$ the parameter representing 
the probability that any detected photon is coming from the background.

In figure \ref{fig:data} we show the joint empirical distribution of the transmitted and received bits on the $\mathbb{X}$ and $\mathbb{Z}$ bases obtained in one run with the best environmental conditions (i.e., with additional background), for the case $p_{\mathbb{Z}}=49 \%$ and $p_{\mathbb{X}}= 51\%$. As expected, in this case the QBER is very low: the main source of errors are imperfections in the waveplates used in the measurement, yielding $\Qall{X}=0.33\%$ and $\Qall{Z}=1.48\%$ on average.

In Figure \ref{fig:key_rate_ber} we show the measured experimental key rates for each data set and for both general and pragmatic secrecy. First of all, let us recall that, in order to consistently compare the secrecy rates obtained with general and pragmatic secrecy, the security parameters $\epssec$ and $\deltasec$ have to be chosen so that $\deltasec = \frac{2}{\ln 2}\epssec^2$.
As a performance reference, we plot the asymptotic {theoretical} bound $r = 1-h_2(\Qall{X})-h_2(\Qall{Z})$, holding in the limit of infinite length keys  (labelled as ``asymptotic'' in Fig. \ref{fig:key_rate_ber}) and  the optimal theoretical bound for $\epssec$-GS keys  (labelled as ``numerically optimized $p_{\mathbb{Z}}$'' in Fig. \ref{fig:key_rate_ber}). 
The experimental key rates are  obtained by the following procedure: for each data set the $n$-bit sifted key $\mathbf{X}$ and the $k$-bit parameter estimation string $\mathbf{Z}$ ($\mathbf{X'}$ and $\mathbf{Z'}$) at Alice's (Bob's) side are obtained by the experiment. The error correction is performed on  $\mathbf{X}$ and $\mathbf{X'}$ by using the Winnow scheme; in particular, the Winnow parameters were chosen so that a maximum of 6 subsequent iterations is allowed with block sizes up to 256 bits. We then performed privacy amplification by compressing the error-free keys by multiplication with a random binary Toeplitz matrix.
The amount of compression depends on $\ell$, the secret key length, given by eq. \eqref{eq:l_gs} and \eqref{eq:l_ps} for general and {pragmatic} security, respectively.
On the other hand, the optimal bound for $\epssec$-GS keys is numerically derived by maximizing the secret key rate $r$ (eq. \eqref{eq:sec_key_rate}, with $\ell$ given by eq. \eqref{eq:l_gs}) over $p_Z$, $\Qtol$ and $Q_{\rm{max}}^{\mathbb{X}}$ for each $n$. 

In the numerical procedure used to find the optimal bound for $\epssec$-GS keys,
since an analytical expression is not available for $L_{\rm{EC}}$ or $\varepsilon_{\rm{rob}}$,  
$L_{\rm{EC}}$ is approximated as $L_{\rm{EC}} = 1.1 \cdot n \cdot h_2(\Qall{X})$ and, similarly, $\varepsilon_{\rm{rob}}$ is replaced by the following upper bound  (see equation A5 of ref. \cite{toma12arx} for details):
\begin{equation}
\varepsilon_{\rm{rob}} \leq \exp \left[-\frac{k(\Qtol-\Qall{Z})^2}{1-2\Qall{Z}}\rm{ln}\left(\frac{1-\Qall{Z}}{\Qall{Z}}\right)\right]
\end{equation}
Experimental values obtained for $\varepsilon_{\rm{rob}}$ show that such bound is rather loose. On the other hand, as $\Qall{X}$ increases, the approximate expression for $L_{\rm{EC}}$ is lower than the average value for the Winnow scheme. As a consequence, the experimental secret key rates may slightly exceed the optimal bound in some low QBER cases, as we can see in fig. \ref{fig:key_rate_ber}a.

As a further comment, we note that, for an asymmetric channel with $\Qall{X} < \Qall{Z}$, using the $\mathbb{Z}$ basis for key encoding and $\mathbb{X}$ for eavesdropper detection provides a higher optimal secret key rate (\ref{eq:sec_key_rate}). However, when the two error rates $\Qall{X}$ and $\Qall{Z}$ have similar values, a minor gain in $r$ is obtained.
For instance, when $n = 10^6$, $\epscor = \epssec = 10^{-10}$, with $\Qall{Z} = 4\%$ and $Q_\mathbb{X} = 2\%$, we can achieve $r = 0.31$; by exchanging the role of $\mathbb{Z}$ and $\mathbb{X}$, $r = 0.33$ can be achieved.

In situations such as satellite quantum communications, the amount of sifted bits is expected
to fluctuate as it depends on the variable channel conditions during the passage. 
From the experimental point of view it is easier to fix the values of $p_{\mathbb{Z}}$ and $p_{\mathbb{X}}$ and accumulate data as long as possible. The value of $p_{\mathbb{X}}$ will constrain the ratio between $k$ and $n$ according to the relation $p_{\mathbb{X}}=\frac{1}{1+\sqrt{k/n}}$. 
In the performed experiments, we thus fixed the value of $p_{\mathbb{Z}}$ and $p_{\mathbb{X}}=1-p_{\mathbb{Z}}$. For each value of the background noise we run different acquisitions with $p_{\mathbb{Z}}$ belonging to the discrete set $\{9\%, 16\%, 28\%, 40\%, 49\%\}$.

\begin{figure}[t]
\includegraphics[width=0.9\textwidth]{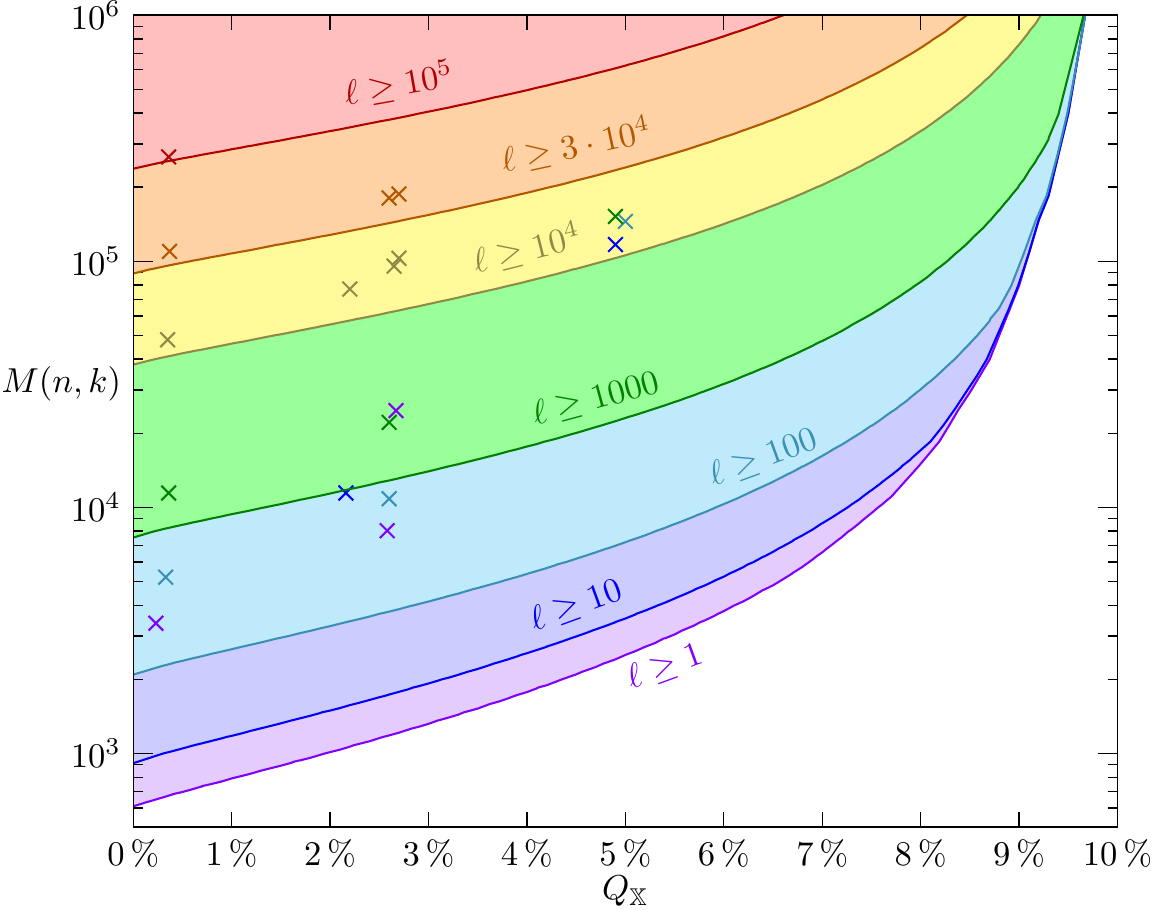}
\caption{{\bf Required bits for a secret key.} {Minimum number of received bits $M(n,k)$ needed to obtain a $\varepsilon_{\rm sec}$-GS key of a given length $\ell$ (as labelled on each curve) versus the quantum BER $\Qall{X}$}. Different colors
divide the regions with different secret key lengths. Crosses represent our experimental results, the colored regions and the solid lines that delimit them are derived from the 
numerically optimized bound, assuming $\Qall{Z} = \Qall{X}$.}
\label{minRecBits}
\end{figure}

Experimental results for the $\epssec$-GS key rates are plotted with thin solid lines, while $\deltasec$-PS key rates are plotted with thin dashed lines; different colors correspond to different $(p_{\mathbb{Z}},p_{\mathbb{X}})$. We used $P_{\rm fail}=10^{-3}$, $\epscor = 10^{-10}$ and $\epssec = 10^{-10}$.
As expected, pragmatic secrecy always allows the achievement of higher secret key rates with respect to general secrecy, which pays the price for the higher level of secrecy it provides.
The gain becomes more evident when the channel becomes noisier and the QBER increases. 
We also observe that with $\Qall{X}=4.9\%$ $\epssec$-GS keys secure are obtained for $p_{\mathbb{Z}}=16\%$, $p_{\mathbb{Z}}=28\%$, {$p_{\mathbb{Z}}=40\%$ and $p_{\mathbb{Z}}=49\%$} and not for $p_{\mathbb{Z}}=9\%$, whereas, when $\Qall{X}=8.3\%$, only keys secure against pragmatic secrecy can be extracted with the parameters we used.

{We point out that the bounds derived for the general and pragmatic secrecy do take into account statistical fluctuations: if the measured $\Qkey{Z}$ is greater than $\Qtol$ the protocol aborts, while for $\Qkey{Z}<\Qtol$ the protocol gives a secure key with security parameter $\epssec$.
As an example, given $Q_X=4.9\%$, $Q_Z=6.0\%$, $n=100000$ and $p_{\mathbb Z}=9\%$, the parameter {$\mu$} which takes into account these fluctuations for general secrecy {(see eq. \eqref{eq:l_gs})}, is approximately equal to 0.15, a value which, for an experimentally realistic number of bits disclosed during the information reconciliation procedure, and even without the contribution of $\Qtol$, yields the impossibility of producing a secret key.
}

Moreover, we notice that higher values of $p_{\mathbb{Z}}$ ($\sim 50\%$) better suit lower values of $n$ for both general and pragmatic secrecy in all considered cases: for instance, when $\Qall{X}=0.3\%$ in the general secrecy case, $p_{\mathbb{Z}}=49\%$ is optimal for 
$n<3\cdot 10^3$; on the other hand, as $n$ increases, it is possible to decrease $p_{\mathbb{Z}}$ and when $n\simeq10^5$ the highest rate is obtained with $p_{\mathbb{Z}}=16\%$. This feature can be understood in the following way: for a short sifted key $\mathbf{X}$, an almost equally long string $\mathbf{Z}$ ($k\sim n$) is needed to reliably detect eavesdropping; when $n$ grows, less bits of $\mathbf{Z}$ (in percentage) are necessary.
In fact, in the large $n$ limit, it is possible to choose $k$ so that $k/n$ vanishes as $n$ goes to infinity and {the secret key rate} approaches the asymptotic bound, $r = 1-h_2(\Qall{X})-h_2(\Qall{Z})$.

It is worth noting that, in the asymptotic limit, a biased choice of the bases gives a higher secure key rate with respect to the BB84
protocol \cite{BB84} whenever $p_{\mathbb{X}}>\sqrt{1/2}$. In fact, in the infinite limit, the fraction of secure over sifted bits is given by $1-2\ent(Q)$ in both cases (for simplicity we here assume $\Qkey{X}=\Qkey{Z}=Q$); however, a biased choice of the bases gives a number of sifted bits that is approximately $p^2_{\mathbb{X}}>1/2$ of the sent bits (also in the finite size regime), 
 while for the BB84 protocol the sifted bits are 1/2 of the sent bits. 
 In particular, by using a large $p_{\mathbb{X}}$, namely $p_{\mathbb{X}}\sim1$, in the infinite key limit we approach a double secret key rate with respect to BB84. 
 In Fig. 2 the asymptotic bound of the secure key rate $r$,  defined as the number of secure bits over number of sent bits, 
 is twice the corresponding asymptotic bound of the BB84 protocol.

With the obtained data we also estimated the minimum number of received qubits $M$ {that are needed} in order to obtain a key of given length $\ell$. In figure \ref{minRecBits} we show this quantity as a function of the QBER (in this case we assumed that $\Qall{X}=\Qall{Z}$).
Solid lines represent the theoretical minimum $M$ necessary to obtain a general secret key for different lengths $\ell$. 
With markers of different colors we indicate the experimental received qubits for the different values of $\ell$. 
Clearly, as the QBER grows, it is necessary to increase the number of exchanged qubits to obtain a given key length $\ell$.
On the other hand, when the channel is almost noiseless, a secret key of reasonable length can be extracted by using a relatively small number of qubits: for instance, more than 1000 secure key bits can be obtained by exchanging less than 20000 photons (see Fig. \ref{minRecBits}).

\section{Discussion}
In conclusion, we have experimentally demonstrated the feasibility of key distillation according to the finite-key analysis proposed in \cite{toma12nco} and compared it with a less stringent definition of security, called pragmatic, that protects the protocol against intercept and resend attacks. 
We compared the two analyses for different amounts of depolarizing noise added to the quantum channel.

With pragmatic security, a significantly secret key rate with finite keys is demonstrated, even in conditions near the theoretical $\Qall{X}$, $\Qall{Z}$ bound of 11\%. Its drawback is the insecurity against collective attacks, which however are not presently available. 
We stress that, when the channel is very noisy ($Q_{\mathbb X}=8.3\%$) no key that is secure against the most general quantum attack could be extracted up to $2\cdot 10^5$ sifted bits; however, by considering only intercept and resend attacks, in this case a secrect key rate up to {7.5\%} was obtained.
When $Q_\mathbb{X}, Q_\mathbb{Z} >11\%$ it is not possible to obtain a secure key even in the asymptotic large $n$ limit.
This shows that, for highly noisy channels, the use of pragmatic secrecy is a viable solution
to obtain some secret bits for a experimentally realistic  number of  exchanged photons.
We believe that our work can have important application for free-space quantum communication and for all QKD scenarios in which the number of exchanged qubits is limited by physical constraints, such as in the inter-satellites link scenario.

\section{Methods}

\subsection{Optical setup}
\label{sect:exp_setup}
\begin{figure*}[t]
\includegraphics[width=1\textwidth]{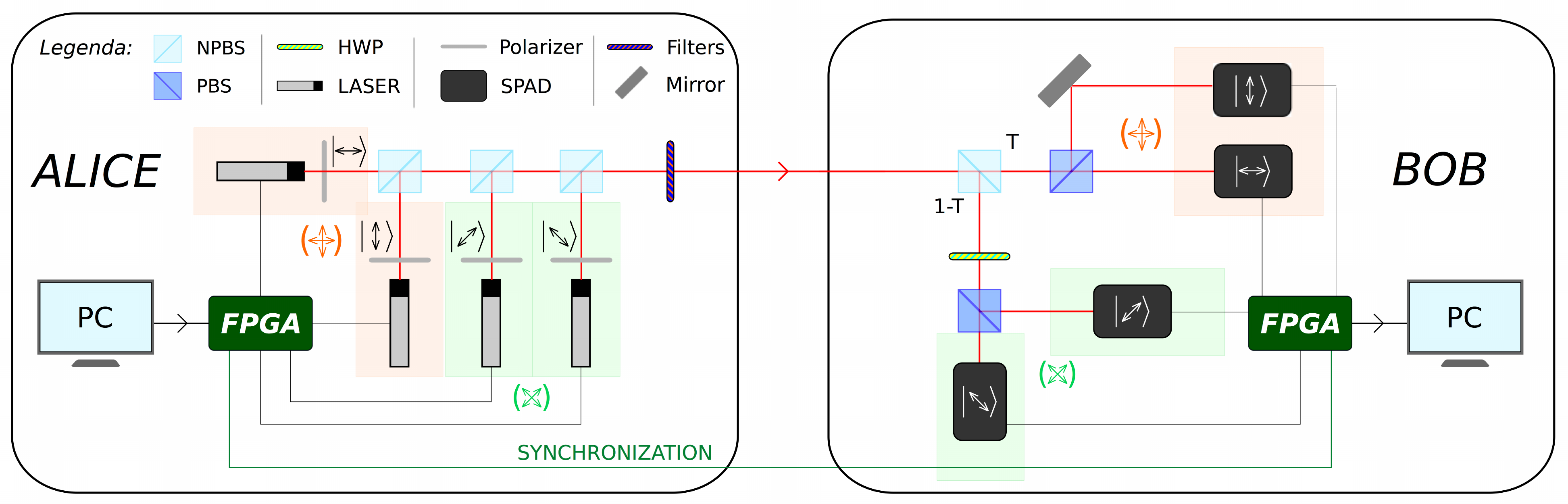} 
\caption{{\bf Schematic of the experimental setup}. The qubits are generated by attenuating four differently polarized lasers.
The FPGA board controls which laser should be turned on in each qubit transmission.
At the receiver side, by a beam splitter with transitivity $T$, Bob perform the measurement in the $\mathbb{X}$ (with probability $T$)
or $\mathbb{Z}$ basis (with probability $1-T$).
NPBS, beam splitter; PBS, polarizing beam splitter; HWP, half wave plate; Filters, 
neutral density filters, SPAD, single photon avalanche diode.}
\label{fig:setup}
\end{figure*}

The optical setup of our prototype implementing the quantum communication is shown in Fig.~\ref{fig:setup}. 
The transmitter (Alice) uses four infrared ($850$nm) attenuated diode lasers driven by a Field Programmable Gate Array (FPGA)
to send the bits $0$ and $1$ encoded in the different polarization bases of the photons. By properly configuring the FPGA, it is possible to
set the probabilities $p_{\mathbb{X}}$ and $p_{\mathbb{Z}}$.
The receiver (Bob) uses a variable  beam splitter (BS) with transmission $T$ to send the received qubits to the measures in the two bases. 
The probability $p_{\mathbb{X}}$ is equal to the transmissivity $T$ of the BS.
On one BS output, a polarizing beam splitter (PBS) and two single photon avalanche photodiodes (SPAD) 
measure the photons in the $\mathbb{X}$ basis; on the other side a half-wave plate (HWP) is positioned before the PBS 
to allow the measurement in the $\mathbb{Z}$ basis. 
The counts detected by the four SPAD are stored on a second FPGA.
A cable between the two FPGA is also used along for synchronization. 

Concerning the transmitted qubits, we used the same data structure 
of a recent free-space QKD implementation~\cite{cana11isa} based on the B92 protocol~\cite{benn92prl}.
A raw key is composed into $N$ packets of $2880$ bits each, which are in turn divided into $12$ frames for the ease of synchronization. In fact, each frame consists of $11$ header slots and $240$ payload slots, each with a duration of $800$ ns. The header exhibits the pattern ''100000xxxx1'', where ''xxxx'' is the $4$-bit frame number, encoded one bit per slot in a pulse-duration modulation of the synchronization beam (a $400$ ns or $200$ ns pulse encode the bit $1$ or $0$, respectively). As regards the payload slots, the first $200$ ns are used to send the synchronization signal; then, Alice waits $200$ ns and sends two bits separated by $200$ ns. 
It is worth noting that the experimental setup of this protocol is very similar to the original BB84: 
the main difference lies in the interpretation of received bits in the two different bases.

\subsection{Classical post-processing}

{After the parameter estimation phase, information reconciliation, error verification and privacy amplification are performed.}
Information reconciliation aims at correcting the discrepancies between $\mathbf{X}$ and $\mathbf{X'}$ that the channel may have introduced, thus allowing Bob to compute an estimate $\mathbf{\hat{X}}$ of $\mathbf{X}$. As a practical solution, we have chosen the Winnow scheme \cite{buttler03} which, by leveraging Hamming codes of different lengths over multiple iterations, allows an adaptive and lowly interactive error correction and represents a good trade-off between the high interactivity required by CASCADE and the low flexibility of LDPC code 
with limited key length.
 
We fix an upper bound $P_{\rm{fail}}$ to the probability of a reconciliation failure and, under this constraint, we optimize the parameters of the Winnow scheme in order to minimize the expected (average) classical information leakage $\expect[L_{\rm{EC}}]$.
First, given the average QBER on the $\mathbb{X}$ basis $\Qall{X}$, a threshold $Q_{\rm{max}}^{\mathbb{X}} > \Qall{X}$ is fixed so that the empirical QBER $\Qkey{X}$ in the sifted key is higher than $Q_{\rm{max}}^{\mathbb{X}}$ with probability less than $P_{\rm{fail}}/2$. 
Then, the block sizes are chosen so that the output BER is lower than $P_{\rm{fail}}/(2n)$ whenever $\Qkey{X} < Q_{\rm{max}}^{\mathbb{X}}$ and $\expect[L_{\rm{EC}}]$ is minimized, as detailed in \cite{cana11isa}.

Subsequently, an error verification mechanism such as the one proposed in \cite{toma12nco} ensures that the protocol is $\epscor$-correct, i.e., that $\prob[\mathbf{X} \neq \mathbf{\hat{X}}] < \epscor$, by comparing hashes of $(\lceil \log_2 (P_{\rm{fail}}/\epscor) \rceil)$ bits. 
Namely, Alice chooses the hash function $g$ randomly and uniformly from a class of universal$_2$ hash functions \cite{carter79} (the class of Toeplitz matrices in our experimental setup) and computes her hash value $g_{\rm{A}} = g(\mathbf{X})$. 
She then sends $g_{\rm{A}}$ and a compact representation of $g$ to Bob, who computes $g_{\rm{B}} = g(\mathbf{\hat{X}})$. The protocol aborts if the two hashes are different, i.e., if $g_{\rm{A}} \neq g_{\rm{B}}$.

Finally, during the so-called privacy amplification,  $\mathbf{X}$ and $\mathbf{\hat{X}}$ are compressed by means of a function which is, again, randomly and uniformly chosen from a class of universal$_2$ hash functions, in order to get the final secret keys $\key$ and $\mathbf{\hat{S}}$.
The length $\ell$ of the final key and the corresponding amount of compression depend on the required level of secrecy, on the overall classical information leakage $L_{\rm{EC}} + \lceil \log_2 (P_{\rm{fail}}/\epscor) \rceil$, on the assumed attacker's model and on the estimate of the information leaked to the eavesdropper during the transmission over the quantum channel.

\subsection{Proof of pragmatic secrecy}

Proof of Theorem 1: let $t$ be the number of qubits observed and measured by Eve on the $\mathbb{X}$ basis among the $n$ sifted bits. 
 Then the R\'enyi entropy of order 2 for the sifted key, given all the information available to the eavesdropper, is lower-bounded by
 \begin{equation}
  R(\mathbf{X}|V) \geq n_{\rm{EC}} - t, \label{eq:renyi}
 \end{equation}
being $R(\mathbf{X}|V) = -\sum_{v} p_V(v)\log_2\left(\sum_\mathbf{s} p ^2_{\key|V}(\mathbf{s}|v)\right)$.

Let us define the following pairs of complementary events, namely: let $A = \{\Qkey{Z} > \Qtol\}$ and $\bar{A} = \{\Qkey{Z} \leq \Qtol\}$ be the aborting and non-aborting events, whereas $R = \{R(\mathbf{X}|V) \geq n_{\rm{EC}} - a\}$ and $\bar{R} = \{R(\mathbf{X}|V) < n_{\rm{EC}} - a\}$ define the events of acceptable and non-acceptable eavesdropping rate, respectively. Then, 
\begin{eqnarray}
\ent(\key|V) &=& \expect[\log_2 \prob(\key|V)|\bar{A}]= \expect[\log_2 p(\key|V)|R, \bar{A}]\prob[R| \bar{A}]+ \expect[\log_2 \prob(\key|V)|\bar{R}, \bar{A}]\prob[\bar{R}| \bar{A}]\,.
\end{eqnarray}

The multiplication of $\ent(\key|V)$ by the probability of not aborting yields
\begin{eqnarray}
\prob[\bar{A}]\ent(\key|V)&=& \expect[\log_2 p(\key|V)|R, \bar{A}]\prob[R, \bar{A}]+ \expect[\log_2 \prob(\key|V)|\bar{R}, \bar{A}]\prob[\bar{R}, \bar{A}]\\
&\leq& \expect[\log_2 p(\key|V)|R, \bar{A}] + \ell\prob[\bar{R}, \bar{A}]\,.
\end{eqnarray}
Finally, by applying corollary 4 in Ref. \cite{bennett95} to a possibly aborting protocol that outputs a $\ell$-bit key (i.e., $\ent(\ukey) = \ell$), we have, for every $a, \ell$,
\begin{eqnarray} 
\label{eq:ileak_bound}
 && \prob[\Qkey{Z} \leq \Qtol](\ell - \ent(\key|V))
 \leq \frac{2^{-(n_{\rm{EC}}-\ell-a)}}{\ln2} + \ell  \prob[R(\mathbf{X}|V) < n_{\rm{EC}} - a, \Qkey{Z} \leq \Qtol]\,.
\end{eqnarray}

From (\ref{eq:renyi}), we can upper bound the probability on the right-hand side of (\ref{eq:ileak_bound}) as 
\begin{eqnarray}
\label{eq:bound_prob_renyi}
\prob[R(\mathbf{X}|V) < n_{\rm{EC}} - a, \Qkey{Z} \leq \Qtol] &\leq& \prob[t > a, \Qkey{Z} \leq \Qtol]\\
\label{eq:bound_prob_renyi_fact}
 &=& \prob[t > a] \prob[\Qkey{Z} \leq \Qtol],
\end{eqnarray}
since the two events in the right-hand side brackets of equation (\ref{eq:bound_prob_renyi}) refer to disjoint qubit sets, namely those encoded in the $\mathbb{X}$ and $\mathbb{Z}$ basis, respectively, and are therefore independent.
Furthermore, according to the selective individual attack model with attack rate $q$, $t$ is a binomial random variable with parameters $(n,q)$. Similarly, the number of measured errors on the $\mathbb{Z}$ basis, $k\Qkey{Z}$ is a binomial random variable with parameters $(k,\Qall{Z})$ and $\Qall{Z} = q/2$.
Therefore, we can rewrite equation  (\ref{eq:bound_prob_renyi_fact}) as
\begin{eqnarray}
\prob[t > a] \prob[\Qkey{Z} \leq \Qtol] &=& (1-F_{n,q}(a)) (F_{k,q/2}(k\Qtol)) \\
& = &  I_q(a+1,n-a) I_{1-q/2}(k(1-\Qtol),k\Qtol+1),
\label{eq:pbin5}
\end{eqnarray}
with $F_{n,q}(\cdot)$ denoting the cumulative distribution function of a binomial random variable with parameters $(n,q)$, and similarly for $F_{k,q/2}(\cdot)$. The last step is then assured by equation 6.6.4 in Ref. \cite{Abramowitz}.    

Eventually, condition (\ref{eq:thm_pragmatic_secrecy}), together with definition (\ref{eq:fdef}) and given that $\prob[\Qkey{Z} \leq \Qtol] = 1- \pabort$, ensures that for any $q \in [0,1]$ we get
\begin{equation} \label{eq:ileak_bound_faq}
  \ell - \ent(\key|V) \leq \frac{\deltasec}{1-\pabort}, \qquad \forall \ a, \ell.
\end{equation}

Relationship between equation  (\ref{eq:PSdef}) and (\ref{eq:GS_sec}): the Pinsker inequality (see section 11.6 in \cite{CoverThomas} and \cite{Wilde}) ensures that
\begin{equation}
\frac{1}{2}\lVert p_{\key V}-u_{\key}q_V \rVert_1 \leq \sqrt{\frac{\ln 2}{2} \kldist(p_{\key V}||u_{\key}q_V)}\,
\end{equation}
where $u_{\key}$ is the uniform distribution on $\key$ and $\kldist(p||q)$ is the relative entropy between the $p$ and $q$ distributions.
By minimizing each term with respect to $q_V$, we get
\begin{eqnarray}
\min_{q_V} \frac{1}{2}\lVert p_{\key V}-u_{\key}q_V \rVert_1 &\leq& \min_{q_V}\sqrt{\frac{\ln 2}{2} \kldist(p_{\key V}||u_{\key}q_V)}\\
&=& \sqrt{\frac{\ln 2}{2} \kldist(p_{\key  V}||u_{\key}p_V)} \label{eq:min_pinsk_1}\\
&=& \sqrt{\frac{\ln 2}{2}(\ent(\ukey)-\ent(\key|V))}\,,
 \label{eq:min_pinsk_2}
\end{eqnarray}
where (\ref{eq:min_pinsk_1}) is due to $\kldist(p_{\key V}||u_{\key}q_V) = \kldist(p_{\key V}||u_{\key}p_V) + \kldist(p_V||q_V) \leq \kldist(p_{\key V}||u_{\key}p_V)$. 
It is then straightforward to see that
\beq
\ent(\ukey)-\ent(\key|V) \leq \frac{2}{\ln 2}\frac{\epssec^2}{1-\pabort} \Rightarrow \min_{q_V} \frac{1}{2}\lVert p_{\key V}-u_{\key}q_V \rVert_1 \leq \frac{\epssec}{(1-\pabort)}.
\eeq

Relationship between equation (\ref{eq:PSdef}) and (\ref{eq:PSdef_bis}): the uniformity condition trivially derives from the fact that $\ent(S|V) \leq \ent(S)$.
Also, from basic information theory, we know that
\begin{equation}
\mutinfo(\key;V) = \ent(\key) - \ent(\key|V) \leq \ent(\ukey) - \ent(\key|V),
\end{equation}
since $\key$ has maximal entropy (i.e., $\ent(\key) = \ell$) if and only if it is uniformly distributed.
Now, since condition (\ref{eq:PSdef}) is verified for any IS attack strategy, and therefore for any outcome $V$ of the eavesdropper measurement on the quantum system $E$, the security condition directly follows.

\section{Acknowledgements}
This work has been carried out within the Strategic-Research-Project QUINTET of the Department of Information Engineering, University of Padova and the Strategic-Research-Project QUANTUMFUTURE of the University of Padova.
Correspondence and requests for materials should be addressed to P.V.: email: paolo.villoresi@dei.unipd.it

\section{Author contributions}
P.V. conceived the work, D.B., G.V. and P.V. designed and performed the experiments, M.C and N. L. analyzed the data and the key extraction, N. L., M. C. and G. V. contributed the secrecy proofs.
All authors discussed the results and contributed to the final manuscript.

\section{Competing financial interests}
The authors declare no competing financial interests.

\end{document}